\documentclass[sigconf, nonacm]{acmart}
\acmConference[EASE 2025]{The 29th International Conference on Evaluation and Assessment in Software Engineering}{17–20 June, 2025}{Istanbul, Turkey}

\usepackage{amsmath,amsfonts}
\usepackage{algorithmic}
\usepackage{graphicx}
\usepackage{textcomp}
\usepackage{xcolor}

\usepackage{epstopdf}
\usepackage{booktabs}
\usepackage{multirow}
\usepackage{url}
\usepackage{flushend}
\usepackage{soul}

\soulregister\cite7 
\soulregister\ref7 
\soulregister\url7 

\graphicspath{{./images/}}

\begin{document}

\title{Towards Effective Issue Assignment using Online Machine Learning}

\author{Athanasios Michailoudis, Themistoklis Diamantopoulos, Antonios Favvas, and Andreas L. Symeonidis}
\email{aamichail@ece.auth.gr, thdiaman@issel.ee.auth.gr, afavvask@ece.auth.gr, symeonid@ece.auth.gr}
\affiliation{%
  \institution{Electrical and Computer Engineering Dept., Aristotle University of Thessaloniki}
  \city{Thessaloniki}
  \country{Greece}
}

\begin{abstract}
Efficient issue assignment in software development relates to faster resolution time, resources optimization, and reduced development effort. To this end, numerous systems have been developed to automate issue assignment, including AI and machine learning approaches. Most of them, however, often solely focus on a posteriori analyses of textual features (e.g.\ issue titles, descriptions), disregarding the temporal characteristics of software development. Thus, they fail to adapt as projects and teams evolve, such cases of team evolution, or project phase shifts (e.g.\ from development to maintenance). To incorporate such cases in the issue assignment process, we propose an Online Machine Learning methodology that adapts to the evolving characteristics of software projects. Our system processes issues as a data stream, dynamically learning from new data and adjusting in real time to changes in team composition and project requirements. We incorporate metadata such as issue descriptions, components and labels and leverage adaptive drift detection mechanisms to identify when model re-evaluation is necessary. Upon assessing our methodology on a set of software projects, we conclude that it can be effective on issue assignment, while meeting the evolving needs of software teams.
\end{abstract}



\maketitle

\section{Introduction}
Contemporary software engineering methodologies have led teams to adopt cloud-based platforms to host their code and manage their development processes.
Code hosting facilities such as GitHub\footnote{\url{https://github.com}} are commonly used to store code, facilitate collaboration between developers and manage version control, while issue tracking systems such as Jira\footnote{\url{https://jira.atlassian.com}} are widely utilized for issue tracking and project management. 
These tools generate a significant amount of data related to software development, which, in turn, can be analyzed to gain insight into workflows, identify issues, and make data-driven decisions to improve productivity and quality.

Numerous approaches aspire to achieve accurate issue assignment from issue report (meta)data \cite{MurphyCubranic:BugTriaging1, Anvik:BugTriaging2, Xuan:BugTriaging3, Ahsan:TopicModeling1, Yang:TopicModeling2, Naguib:TopicModeling3, Xia:TopicModeling4, Diamantopoulos:TopicModeling5, Mani:Embeddings1, Lee:Embeddings2, Jonsson:Ensemble}, others have focused on issue prioritization \cite{Sharma:Priority1, Tian:Priority2, KanwalMaqbool:Priority3} and/or bug severity estimation \cite{Diamantopoulos:TaskImportancePrediction, Lamkanfi:CoarsegrainedSeverity1, Yang:CoarsegrainedSeverity3}, while some have even attempted developer role inference \cite{Li:Contributions1, Onoue:Contributions2, Gousios:DeveloperContributions, Lima:DeveloperContributions, Papamichail:ContributorRoles}.
However, although effective, they also suffer from limitations.
Specifically, most approaches conduct a posteriori analyses solely on textual features of issues, hence omitting important information about project characteristics (i.e.\ components, labels), and their interconnectivity. 
Furthermore, since software teams and projects are dynamic entities, with team changes, project shifts, priority changes, etc., in vitro experiments fail to account for the temporal characteristics of the task assignment process, thus leading to inaccurate estimations over time.

In this work, we propose a methodology that overcomes these limitations.
Specifically, we design a system that takes into account the state of the project as a whole, considering the contributors, the issue titles, descriptions, and metadata, while also leveraging temporal elements of software projects, by modeling the issues of each project as data streams (i.e.\ one issue appearing after the other). To mine these data, we employ online machine learning algorithms that are able to efficiently model possible drifts in software projects (e.g.\ contributor shifts, context switches).
Our approach is further extended by designing a model ensemble with drift detection capabilities, which supports efficient re-training for continuous issue assignment adaptation.

The remainder of the paper is structured as follows: Section \ref{sec:relatedwork} reviews previous research in the areas of automated issue assignment, while  Section \ref{sec:methodology} presents our methodology for assigning issues to each developer, according to the issue's metadata and the project's temporal state.
Section \ref{sec:evaluation} evaluates our methodology using a set of software projects extracted from a Jira installation, Section \ref{sec:discussion} discusses possible limitations of our approach, and finally Section \ref{sec:conclusion} discusses our conclusions and provides ideas for future work.

\section{Related work} \label{sec:relatedwork}
Several approaches have been proposed in recent years to achieve automated issue assignment, ranging from topic modeling and developer ranking, to deep learning approaches and LLMs. 

Early approaches treated bug assignment as a text classification problem.
In specific, Cubranic \& Murphy \cite{MurphyCubranic:BugTriaging1} developed a supervised Bayesian model, trained on bug reports from Eclipse and attempted to predict assignees, based on the report's title and description.
Anvik et al. \cite{Anvik:BugTriaging2} improved upon prior work by including metadata, such as issue severity and priority.
The study also introduced an SVM classifier, found to perform better than the Bayesian model.
Furthermore, the researchers applied developer activity filtering, removing inactive or inexperienced developers (with respect to the project). 
Finally, Xuan et al. \cite{Xuan:BugTriaging3} combined a Naive Bayes classifier along with Expectation-Maximization to create a Weighted Recommendation List, allowing multiple developers to be probabilistically assigned to a bug report.
Though innovative, the method underperformed against prior works in terms of accuracy.

Later approaches opted for more sophisticated techniques, such as topic modeling. Ahsan et al. \cite{Ahsan:TopicModeling1} expanded on feature selection; along with report ``summary'' and ``description'' they also included the title, product name, and operating system relevant to the bug.
Moreover, they introduced Latent Semantic Indexing (LSI) to eliminate noise and preserve useful semantic relationships.
Lastly, they conducted a comparative study between 7 different supervised-ML algorithms, from which the SVM yielded the best results.
Naguib et al. \cite{Naguib:TopicModeling3} created developer activity profiles to exploit for assignee recommendation and suitability ranking, based on older, resolved issues.
Yang et al. \cite{Yang:TopicModeling2} introduced social network analysis between developers and conducted data preprocessing by combining topic modeling (LDA) along with multi-feature extraction (i.e.\ product, component, priority, severity), while a KNN algorithm handled severity-prediction. Ultimately, the approach outperformed traditional methods in both bug-triaging and severity-prediction.

Recent efforts leverage word embedding techniques (i.e.\ Word2Vec \cite{Guo:word2vec}, GloVe \cite{Pennington:Glove}), to capture bug report semantics.
Lee et al. \cite{Lee:Embeddings2} combined Word2Vec with Convolutional Neural Networks (CNNs) to process bug reports from industrial projects, thus addressing challenges such as multilingual bug reports and domain-specific jargon.
The CNN outperformed traditional machine learning models and human triagers in both accuracy and efficiency, particularly in large industrial settings.
On the other hand, Mani et al. \cite{Mani:Embeddings1} developed ``DeepTriage'', a Deep Bidirectional Recurrent Neural Network (DBRNN-A), that paired word embeddings with an attention mechanism, allowing it to focus on the most critical parts of the bug report.
Their research also demonstrated the effectiveness of cross-project transfer learning, where models trained on one project could be applied to other projects with minimal performance loss. 

Although the approaches described above are effective in certain scenarios, most of them rely on traditional batch learning methods, which assume static datasets with immutable underlying data distribution.
However, nowadays, software engineering data are characterized by velocity and variability. Rapid evolution on underlying characteristics of software projects, such as developer availability and expertise, and project priorities may often fluctuate and affect issue assignment.

To address these challenges, we propose an online automated issue assignment system that continuously adapts to the changes of the software project.
Our system employs \textit{Online Machine Learning} and treats data as a \textit{data stream} rather than static datasets.
It makes predictions and trains concurrently on every data instance, while also detecting and adapting drifts, such as the addition of new developers or the retirement of existing ones from the project.
This design achieves accurate assignment of incoming issues to the most suitable developer by ensuring continuous alignment with the evolving structure and needs of the development team.

\begin{figure*}[!htbp]
\centerline{\includegraphics[width=\textwidth]{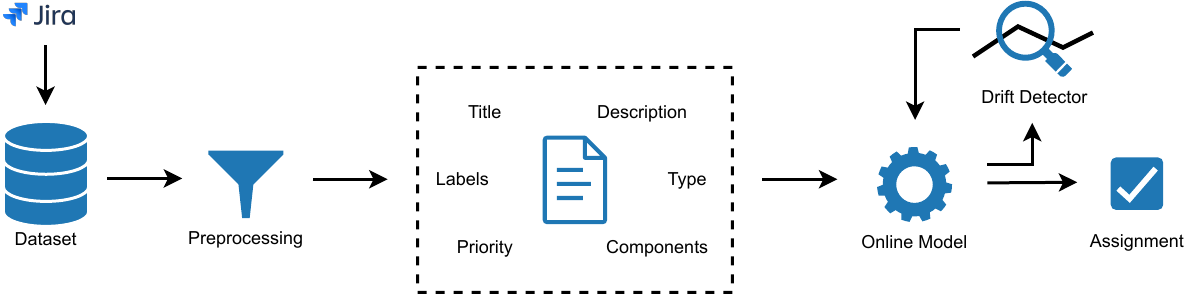}}
\caption{Architecture of our Online Automated Issue Assignment System}\label{fig:system}
\Description{System Architecture}
\end{figure*}

\section{Methodology} \label{sec:methodology}
The architecture of our system is depicted in Figure \ref{fig:system}. We design our methodology using as input a set of Jira projects, from the Jira dataset created by Diamantopoulos et al. \cite{Diamantopoulos:JiraDataset}, which includes more than 1M issues on 656 Apache projects. Furthermore, since our approach leverages online learning, we chose the \textit{River} framework \cite{Montiel:River}, which is a Python library that provides the necessary preprocessing tools and machine learning algorithms.

Upon preprocessing and temporal alignment (subsection \ref{sec:prep}), we feed all data instances into our system, one-at-a-time, in chronological order.
Our online model (subsection \ref{sec:issue_assignement}) then predicts the most suitable developer for the current issue (\textit{predict step}), and is immediately updated based on the prediction's validity (\textit{learn/train step}).
Meanwhile, the drift detector (subsection \ref{sec:drifts}) monitors our model's performance to determine fluctuations in the nature of data; should any be detected, corrective measures are taken to ensure accurate and consistent performance at all times.

\subsection{Data Preprocessing} \label{sec:prep}
Since new reports are created constantly, there always is a continuous flow of data, effectively replacing static datasets with dynamic \textit{data streams}.
To simulate data streams, we initially order all retrieved issues of each project in chronological order -from oldest to most recent- according to the issue creation timestamp.
 
Regarding feature extraction, previous work suggests that textual features, such as titles/summaries and descriptions of issue reports, contribute the most in modeling developers' technical skills \cite{MurphyCubranic:BugTriaging1, Ahsan:TopicModeling1}. 
Therefore, we transform them into text vectors through \textit{TF-IDF}, since it is a fast technique that can be easily applied in an online manner. 
However, since we cannot have access to whole datasets to calculate the exact parameters of the vectorizer, the traditional method is slightly modified so that document frequencies and, subsequently, the vector values are updated incrementally, with each new data point.
Furthermore, we also added issue type, as well as relevant labels and components, to extract even more useful information from dependencies \cite{Yang:TopicModeling2}.
As for our ordinal features, we selected issue priority as prior research highlight its relevance for developer assignment based on bug severity \cite{Anvik:BugTriaging2}.

\subsection{Issue Assignment} \label{sec:issue_assignement}
We opted for the Multinomial Naive Bayes model, which is widely used in this domain \cite{MurphyCubranic:BugTriaging1, Xuan:BugTriaging3}.
It allows for incremental updates of model parameters, as every word is assumed to independently contribute to the final classification; thus being ideal for data streams.
It works particularly well on multinomial data distributions such as text data, making it an intuitive fit for TF-IDF representations.
Finally, it has relatively low memory and computational demands. 

To reinforce our model's performance and resilience to class imbalance -a common issue in bug triaging- we employ Ensemble Learning \cite{Jonsson:Ensemble}.
We chose AdaBoost \cite{Oza:Adaboost}, which combines multiple ``weak'' learners (e.g., Naive Bayes), to form a ``stronger'' classifier, through appropriate model weighting.
Furthermore, it focuses on misclassified instances by adjusting weights for each observation with each new iteration, emphasizing ``hard-to-classify'' cases in subsequent updates.
We use 10 Naive Bayes models as weak learners to achieve balance between model accuracy and efficiency. 

\subsection{Drift Detection} \label{sec:drifts}
By handling projects as data streams, we can also identify drifts in their distribution. We distinguish among two types of drifts, \textit{concept} and \textit{data} drifts.
\textit{Concept drifts} refer to the shift of correlation between input features and the target variables overtime, meaning that an input results in alternative outputs at different points in time.
For example, the retirement of a domain-expert developer may cause a drop in the system's performance, as it continues assigning new relevant issues to them, despite their absence.
Conversely, \textit{data drifts} refer to shifts of the input data's statistical properties over time.
As a bug triaging example, consider a shift in the types of reported issues due to changes in the software product or user base. The triage model, initially trained on certain keywords or patterns, may become less accurate as the nature of bug reports evolves. 

\begin{figure}[!htbp]
\centerline{\includegraphics[width=\columnwidth]{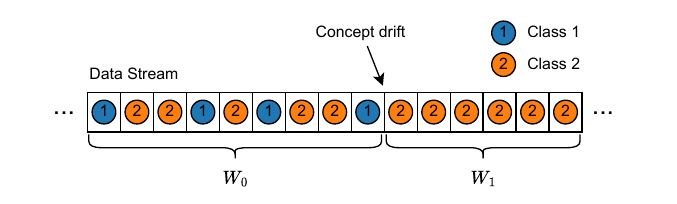}}
\caption{ADWIN drift detection example}\label{fig:adwinexample}
\Description{Data drift example}
\end{figure}

To combat these phenomena and ensure robust system performance, we employ the ADWIN (Adaptive Windowing) \cite{Bifet:ADWIN} drift detector.
It operates by maintaining a variable-length, sliding window, $\textbf{\textit{W}}$, of the most recent data points, and continuously monitoring for changes in data distribution.
Incoming data points are added to \textbf{$\textbf{\textit{W}}$}, while older instances are removed if significant difference is detected between more recent and older data.
This is done by splitting $\textit{\textbf{W}}$ into two sub-windows; \(\textbf{\textit{W}}_0\) for older data ("tail") and \(\textbf{\textit{W}}_1\) for more recent data ("head").
If their mean difference exceeds a certain threshold (calculated by Hoeffding's inequality), ADWIN identifies a drift. At this point, \(\textbf{\textit{W}}_0\) is discarded and \(\textbf{\textit{W}}\) is reset to \(\textbf{\textit{W}}_1\), effectively ensuring that the model does not rely on outdated information.
Figure \ref{fig:adwinexample} depicts an example window of a data stream monitored by ADWIN, where the squares represent sequential input data, that correspond to two distinct classes.

At first, incoming data show a balanced distribution between the two classes; however, the balance shifts towards Class 2, as the issues progress.
ADWIN splits the initial window into windows \(\textbf{\textit{W}}_0\) and \(\textbf{\textit{W}}_1\), and monitors the difference in mean assignment rates between them. The mathematical expression that describes that comparison is:
\[\left|\mu_{W_{0}} - \mu_{W_{1}}\right|\geq \varepsilon_{cut},\]
where $\mu_{W_{0}}$ and $\mu_{W_{1}}$ are the average assignment rates for each window, and $\varepsilon_{cut}$ is a predefined threshold.
When mean difference exceeds that threshold, ADWIN identifies a concept drift.

ADWIN was chosen based on its suitability for online learning and data streams, since it was specifically designed for real-time updates and low computational demands. 
It is also highly versatile, as it can be used to detect both types of drifts, and is easily employable out-of-the-box, since no manual tuning is required.

\begin{figure*}[!ht]
    \centerline{\includegraphics[width=\textwidth]{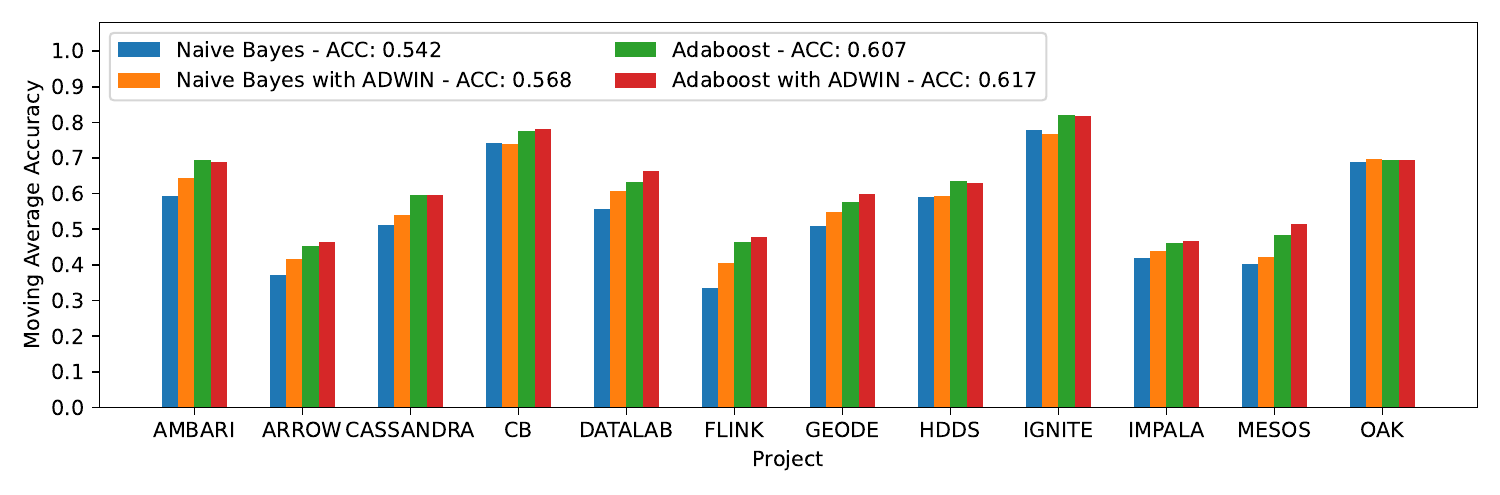}}
    \caption{Comparison of our model (Adaboost with ADWIN) vs 3 baselines, Naive Bayes, Naive Bayes with ADWIN, Adaboost}
    \Description{Comparison of models}
    \label{fig:naivebayes_vs_adaboost}
\end{figure*}

For our problem, we set up ADWIN to monitor our model's performance and detect drifts causing accuracy drops. Should a drift is detected, we identify and replace AdaBoost's five lowest-performing individual classifiers with new ones.
The reasoning behind retaining the strongest-performing classifiers was to avoid complete retraining of the system, which resulted in significant performance drop with every drift detection.
Furthermore, we consider developer inactivity; upon evaluation of our model's prediction, we update an index with all developers that actually resolved an issue during the last 100 issues, and assign them a weight of 1.
If a developer has not resolved any issues in that interval, they are assigned a weight of 0, effectively eliminating their assignment probability until they appear again in the index.
This method assumes that long periods of assignee inactivity might suggest (temporary) unavailability, re-assignment or even retirement from the project.  

\section{Evaluation} \label{sec:evaluation}
\subsection{Evaluation Framework} \label{sec: projects}
We select projects with 5 or more assignees, who were assigned at least 80 issues each during the project.
We retrieve issues that include the fields for our training (see subsection \ref{sec:prep}) and an assignee.
Thus, we end up with 12 projects, shown in Table \ref{tab:project_stats}. 

\begin{table}[!ht]
    \caption{Project Statistics for the Dataset Projects}
    \label{tab:project_stats}
    \centering
    \begin{tabular}{lccc}
        \toprule
        Project & \#Assignees & \#Issues & \#Issues per Assignee \\
        \midrule
        AMBARI & 8 & 1131 & 141 \\
        ARROW & 17 & 4210 & 247 \\
        CASSANDRA & 8 & 909 & 113 \\
        CB & 5 & 1007 & 201 \\
        DATALAB & 8 & 1405 & 175 \\
        FLINK & 29 & 4506 & 155 \\
        GEODE & 7 & 928 & 132 \\
        HDDS & 8 & 1498 & 187 \\
        IGNITE & 5 & 804 & 160 \\
        IMPALA & 12 & 2025 & 168 \\
        MESOS & 14 & 1597 & 114 \\
        OAK & 5 & 765 & 153 \\
        \bottomrule
    \end{tabular}
\end{table}

To evaluate our online model's performance, we monitor its \textit{Moving Average Accuracy}, a metric computed by \textit{River} \cite{Montiel:River} as the aggregate accuracy of all predictions up to the current point (so it is updated with each new prediction). 
We compare our model against three baseline methods. 
Those are Naive Bayes, a pipeline combining Naive Bayes with ADWIN where drift detection triggers retraining, and ensemble learning through AdaBoost. 
Finally, we add our proposed method, i.e.\ a pipeline consisting of AdaBoost and ADWIN that considers issue resolution frequencies and replaces the lowest-performing half of base models upon drift detection. 

\subsection{Online Classification Evaluation}
Figure \ref{fig:naivebayes_vs_adaboost} presents an overall comparison between the tested models.
As is evident, using Drift Detectors in both Naive Bayes and AdaBoost proves to be beneficial -or at least as effective- suggesting that ADWIN improves accuracy by helping the models adapt to drifts.
As expected, the baseline model struggles the most, though its performance implies that Online Learning is a valid approach to the task.
The graph also indicates that the introduction of ensemble learning (green bar) provides the most benefit to overall performance.
In most projects, our method (AdaBoost with ADWIN) outperforms Naive Bayes with ADWIN, highlighting AdaBoost's high effectiveness when combined with ADWIN's drift detection. 

The accuracy of the model varies across projects, which is related to the number of developers; the more developers that are occupied on a project, the harder the issue assignment.
In spite of that, mean accuracy settles on ~62\% across all projects, while performance never drops below 46\% -for ARROW and FLINK, two Apache projects with the most assignees- and it even goes as high as close to 80\% -a number comparable to earlier work in the field \cite{Jonsson:Ensemble,Diamantopoulos:JiraDataset}.

\subsection{Project Analysis}
To further examine our model's training process, we provide a detailed case study on Datalab; a platform for creating exploratory data science environments.
Though a typical project of our dataset, conclusions for all other projects are similar.

\begin{figure*}[!t]
    \centerline{\includegraphics[width=\textwidth]{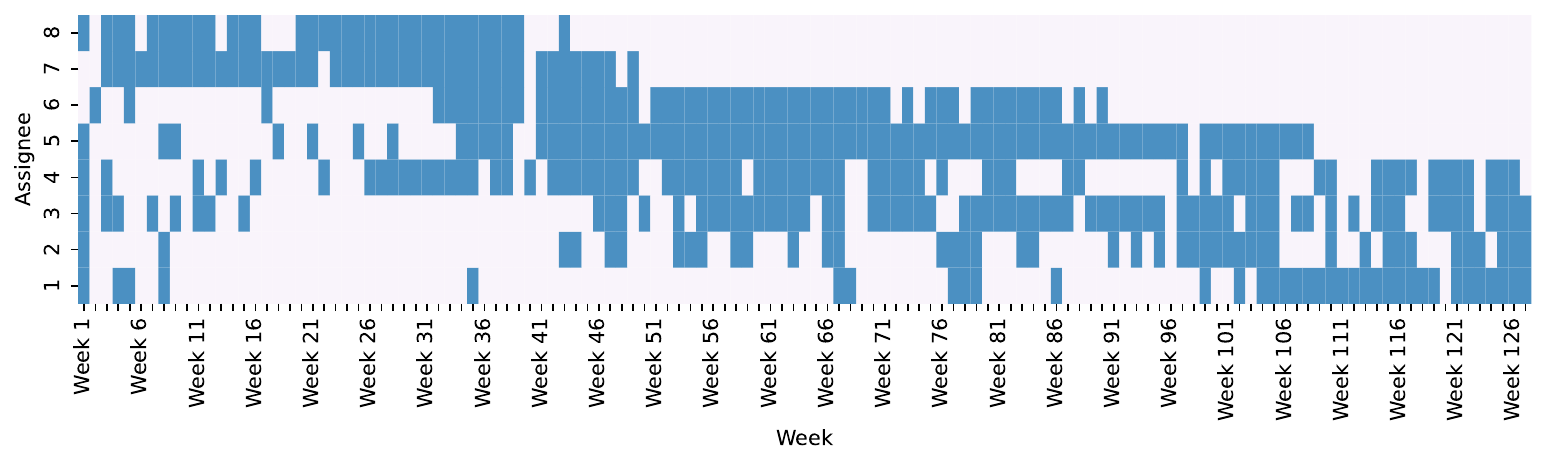}}
    \caption{Developer Activity during Datalab Project}
    \Description{Developer Activity}
    \label{fig:activity}
\end{figure*}

As a first step of our analysis, we examine the (in)activity periods of developers, since not all developers are always available or receive the same amount of workload at all times during the course of a project. Some may be inactive for long periods or even leave the project, so they will not be available to be assigned issues that otherwise fit their profile.
On the other hand, some may be more suited to resolve specific bugs; hence, class imbalances occur regarding the number of issues assigned to each developer. 
Figure \ref{fig:activity} depicts the developers' weekly activity.
Each developer is considered active in a week if he/she has been assigned at least one issue. 
For instance, Assignees 7 and 8 were only active for approximately the first third of the project, whereas Assignee 1 was sparsely taking on issues up until the last quarter of the project.
These insights were also the reason ensemble learning was selected (\ref{sec:issue_assignement}), as it prevents our model from fixating on a specific ``experienced'' developer.

\begin{figure}[ht]
    \centerline{\includegraphics[width=\columnwidth]{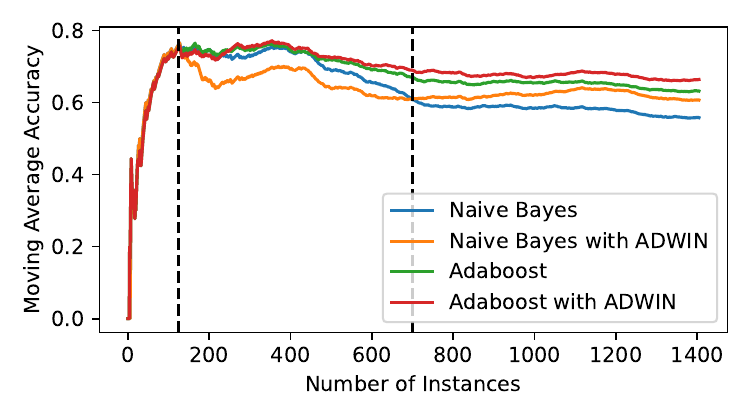}}
    \caption{Moving Average Accuracy of the four configurations (three baselines and our proposed model) for project Datalab}
    \Description{Datalab Performance Evolution}
    \label{fig:issuesexample}
\end{figure}

In Figure \ref{fig:issuesexample} we present a comparison among the four different models with respect to the evolution of their average accuracy over the course of project Datalab (i.e.~computed as an aggregate up to the number of evaluated instances).
As expected, all models begin with low accuracy, but rapidly improve early on.
Then, a drift is detected (black dashed line), right around 160 data instances.
Interestingly, at that point Naive Bayes complemented by ADWIN experiences a quick drop in accuracy, possibly since the whole model is being reset and has to be retrained.
Conversely, the Naive Bayes model seems more stable initially; however, it lags behind the rest of the methods at a later stage, as it is incapable of adapting over drifts, resulting in overall mean accuracy less than 60\%. 

The conclusions for ensemble learning are similar.
Since our method forces AdaBoost to drop half of the base models, it experiences a small drop in its performance.
However, it quickly recovers and surpasses plain AdaBoost, peaking slightly above 0.7.
After around 300 instances, both models' accuracies stabilize, consistently showing better performance than Naive Bayes models.
Our model steadily maintains higher accuracy, even after a second drift is detected, at around 700 instances, indicating that our method is stable and effectively adapts to shifts in data distribution.

\section{Discussion} \label{sec:discussion}
Upon demonstrating our case for applying online learning on issue assignment, we further discuss the scope of our approach as well as any open issues to be addressed in future research in this domain.

Concerning dataset selection, open-source Apache Software Foundation projects may raise applicability concerns of our method, to a broader spectrum of software projects. Most existing approaches primarily focus on bug triaging \cite{MurphyCubranic:BugTriaging1, Anvik:BugTriaging2, Xuan:BugTriaging3, Guo:word2vec, KanwalMaqbool:Priority3, Lee:Embeddings2, Mani:Embeddings1, Ahsan:TopicModeling1}, which includes bug-tracking datasets and is specific to the maintenance phase of the software development process.
Also, our filtering (Section \ref{sec: projects}) retrieved more structured projects, involving smaller, active teams.
To that we argue that these prerequisites follow current practices in open-source projects, where core development is handled by few specific individuals (``maintainers``), with external contributions coming from a larger community.
Also, the diversity of the selected projects, supported by our model's performance, suggests adequate applicability to a broader spectrum of tasks. 

Regarding text preprocessing, we rely on TF-IDF vectorization, instead of word embeddings, which are typically used in similar work \cite{Mani:Embeddings1, Lee:Embeddings2}.
Though a more sophisticated method and initially explored, our experiments showed no significant difference of the two on our dataset.
Lack of strong semantic connotations between terms paired with TF-IDF's suitability for online tasks explain our decision, although the extensive use of embeddings in current literature, prompts us to investigate possible value in their use.

Considering the selected framework, \textit{River} \cite{Montiel:River} provides a variety of tools to tackle online tasks.
However, it lacks robust implementations (i.e.\ SVMs, Neural Networks) appearing regularly in bug triage problems \cite{Anvik:BugTriaging2, Ahsan:TopicModeling1, Naguib:TopicModeling3, Matsoukas:Commits, Mani:Embeddings1, Lee:Embeddings2}.
To compensate and enhance our system's performance we introduced ensembles \cite{Jonsson:Ensemble}, consisting of a Bayesian model, as suggested in relevant literature \cite{MurphyCubranic:BugTriaging1, Xuan:BugTriaging3}.
Evidently, our results already demonstrate the benefits of an online approach, even with ``weaker'' models, as it mitigates limitations of traditional batch learning (i.e.\ class imbalance), while reserving experimentation with more models (e.g.\ Random Forests) and/or frameworks (e.g.\ CapyMOA\footnote{\url{https://capymoa.org}}) for future work.

Finally, and most importantly, we explore an online approach that effectively handles the issues of software project as data streams, whereas current literature mainly focuses on traditional machine learning implementations.
Due to the aforementioned limitations, a direct comparison with other automated issue assignment approaches was not feasible.
Designing a comparison setting would be an interesting idea for future work.
For now, to further facilitate researchers working in this direction, all data and scripts required to reproduce our findings are available in the following repository: \url{https://github.com/AuthEceSoftEng/online-issues}

\section{Conclusion} \label{sec:conclusion}
Although automated issue assignment is a widely researched challenge, conventional methods often fail to consider project dynamics, such as changing team structures, varying developer availability, and shifts in project focus.
These limitations reduce the effectiveness of current approaches, usually leading to inefficient task allocation and longer resolution times.
In this work, we address these challenges by introducing an online learning approach that continuously adapts to the evolving nature of software projects. 
By processing incoming issues as data streams and exploiting issue metadata, we leverage online ensembles paired with drift detection, to ensure alignment with project conditions; thus, improving both assignment accuracy and responsiveness to evolving project needs.

As future work, we intend to enrich our feature selection, by incorporating more developer-related information (e.g. team communications, involvement in other projects), and also explore the concept of developer inactivity by investigating the duration of a developer's absence before he/she is considered unavailable for issue assignment.
Additionally, we plan to address data drifts by applying ADWIN directly on the features, thus identifying shifts that may cause performance degradation.
Finally, we aim to experiment with other online models that adapt to drifts, such as Hoeffding Adaptive Trees \cite{BifetGavalda:HoeffdingAdaptiveTree} or Adaptive Random Forests \cite{Gomes:AdaptiveRandomForest}.

\section*{Acknowledgment}
Parts of this work have been supported by the Horizon Europe project ECO-READY (Grant Agreement No 101084201), funded by the European Union.

\balance
\bibliographystyle{ACM-Reference-Format}
\bibliography{paper}

\end{document}